\documentclass[%
reprint,
 amsmath,amssymb,
 aps,
]{revtex4-1}

\usepackage{graphicx}
\usepackage{dcolumn}
\usepackage{bm}

\begin{document}

\preprint{APS/123-QED}

\title{Interparticle friction leads to non-monotonic flow curves \\ and hysteresis in viscous suspensions}

\author{Hugo Perrin$^{1,2}$}
\author{C\'ecile Clavaud$^{2}$}%
\author{Matthieu Wyart$^{1}$}%
\author{Bloen Metzger$^{2}$}%
\author{Yo\"el Forterre$^{2}$}%

\affiliation{
$^1$Institute of Physics, Ecole Polytechnique F\'ed\'erale de Lausanne, CH-1015 Lausanne, Switzerland \\
$^2$Aix Marseille Univ., CNRS, Institut Universitaire des Syst\`emes Thermiques et Industriels, 13453 Marseille, France
}
%
%
%


\date{\today}

\begin{abstract}
Hysteresis is a major feature of the solid-liquid transition in granular materials. This property, by allowing metastable states, can potentially yield catastrophic phenomena such as earthquakes or aerial landslides. The origin of hysteresis in granular flows is still debated. However, most mechanisms put forward so far rely on the presence of inertia at the particle level. In this paper, we study the avalanche dynamics of non-Brownian suspensions in slowly rotating drums and reveal large hysteresis of the avalanche angle even in the absence of inertia. By using micro-silica particles whose interparticle friction coefficient can be turned off, we show that microscopic friction, conversely to inertia, is key to triggering hysteresis in granular suspensions. To understand this link between friction and hysteresis, we use the rotating drum as a rheometer to extract  the suspension rheology close to the flow onset for both frictional and frictionless suspensions.  This analysis shows that the flow rule for frictionless particles is monotonous and follows a power law of exponent $\alpha \!= \! 0.37 \pm 0.05$, in close agreement with the previous theoretical prediction, $\alpha\!=\! 0.35$. By contrast, the flow rule for frictional particles suggests a velocity-weakening behavior, thereby explaining the flow instability and the emergence of hysteresis. These findings show that hysteresis can also occur in particulate media  without inertia, questioning the intimate nature of this phenomenon. By highlighting the role of microscopic friction, our results may be of interest in the geophysical context to understand the failure mechanism at the origin of undersea landslides.
\end{abstract}

\maketitle


\section{\label{sec:level1}Introduction}

Particulate media like dry granular materials and suspensions are ubiquitous in geophysical and industrial flows \cite{andreotti2013granular}. Yet, understanding their flowing behavior still remains an important challenge, especially for very concentrated media where these materials can jam and exhibit a liquid to solid transition. For long, dry granular materials and dense suspensions were studied separately and described with very different frameworks as in suspensions, hydrodynamic interactions were believed to prevent physical contacts between particles \cite{mewis2012colloidal}. However, this view is changing as a growing body of evidence now shows that solid contacts and interparticle friction play a major role \cite{guazzelli2018rheology}, controlling for instance the packing fraction and the shear-to-normal stress ratio at which suspensions jam \cite{boyer2011unifying}, the scaling law of the suspension viscosity near jamming \cite{degiuli2015unified} or the dramatic shear-thickening observed in colloidal suspensions when particles interact through an additional short-range repulsive force  \cite{mari2014shear,wyart2014discontinuous,Clavaud2017}.  As a result, a unified description of dry granular flows and dense suspensions has emerged over recent years,  based on a frictional rheology and a pressure-imposed framework \cite{boyer2011unifying,degiuli2015unified,guy2015towards,wang2015constant}.

This analogy between dry granular materials and suspensions has so far mainly focused on their steady flowing behavior. Much less is known about whether such an analogy can be made for the transition between the static and the flowing regimes.  A major feature of the solid to liquid transition in dry granular materials is its hysteretic nature \cite{forterre2008flows}.  When a static granular material is loaded under an imposed shear stress, the level of stress required to trigger the flow is larger than the critical stress below which the flow stops. In gravity-driven flows, like heap flows or flows down inclined planes, this implies that flow starts at an angle $\theta_{\rm start}$ larger than that at which it stops $\theta_{\rm stop}$ \cite{carrigy1970experiments,daerr1999two,pouliquen2002friction}. Since in landslides, the mobilized mass closely relates to the difference between the starting and stopping angles, such hysteretical behavior plays an important role in catastrophic geophysical events. Moreover, when the flow starts at an angle above the stopping threshold, the avalanche quickly accelerates to a finite velocity, eventually causing catastrophic failure. Such velocity-weakening dynamics is observed in many contexts from solid friction \cite{baumberger2006solid} to rupture of granular gouges \cite{leeman2016laboratory} and large landslides \cite{lucas2014frictional}.  Interestingly, catastrophic landslides are also observed for immersed sediments \cite{hampton1996submarine}, where they are recognized as a potential source of tsunamis \cite{Lee2005, lovholt2017}. In such a context, it appears particularly important to know whether the hysteresis observed for dry granular avalanches also occurs when particles are immersed in a fluid and through this question, address whether here too, the solid to liquid transition in dry granular flows and dense suspensions presents similarities.

The origin of hysteresis in particulate media is still debated but several theoretical approaches have put forward inertia as a key ingredient of the mechanism \cite{quartier2000dynamics,DeGiuliPNAS}. For long, this view was supported by the pioneering experiments of Courrech du Pont et al \cite{Courrech}. They studied aerial and submarine granular avalanches in slowly rotating drums using different particle sizes and fluid viscosities. For large inertia characterized by the Stokes numbers  ${\rm St}=\sqrt{\rho_p\Delta \rho gd^3}/18\eta$, where $\rho_p$ is the particle density, $\Delta \rho=\rho_p-\rho_f$ the density difference between the particles and the fluid, $d$ the particle diameter, $g$ the gravity and $\eta$ the fluid viscosity, they reported a large and roughly constant hysteresis of the avalanche angle. However, below a critical Stokes number, hysteresis was shown to  decrease as inertia decreased in the system, suggesting that no hysteresis occurs in fully overdamped suspensions. 

In this article,  we study the avalanches dynamics of non-Brownian granular suspensions in rotating drums for Stokes numbers much smaller than those studied in \cite{Courrech}. Surprisingly, we find a large hysteresis of the avalanche angle even when inertia is negligible. Using a suspension of silica micro-beads where interparticle friction can be turned on or off by screening their short-range repulsive force, we show that microscopic friction is essential to trigger hysteretic avalanches. Finally, examination of the avalanche dynamics allows us to extract the effective frictional rheology of the suspensions. This analysis shows that the hysteresis observed with frictional suspensions arises from a non-monotonic effective friction law with a velocity-weakening regime close to the flow onset. By contrast, in frictionless suspensions, the flow rule is monotonous. Overall, our work reveals the existence of hysteresis and of a velocity-weakening rheology in overdamped frictional suspensions, further unifying the flowing properties of dry granular flows and dense suspensions. Our results show that inertia is not required to observe large hysteresis in granular suspensions but that interparticle friction is key, questioning the  origin of hysteresis in particulate media.

\section{\label{sec:level2}Results}
\subsection{Evidence of hysteresis in overdamped suspensions}

\begin{figure}
\includegraphics[width=0.95\linewidth]{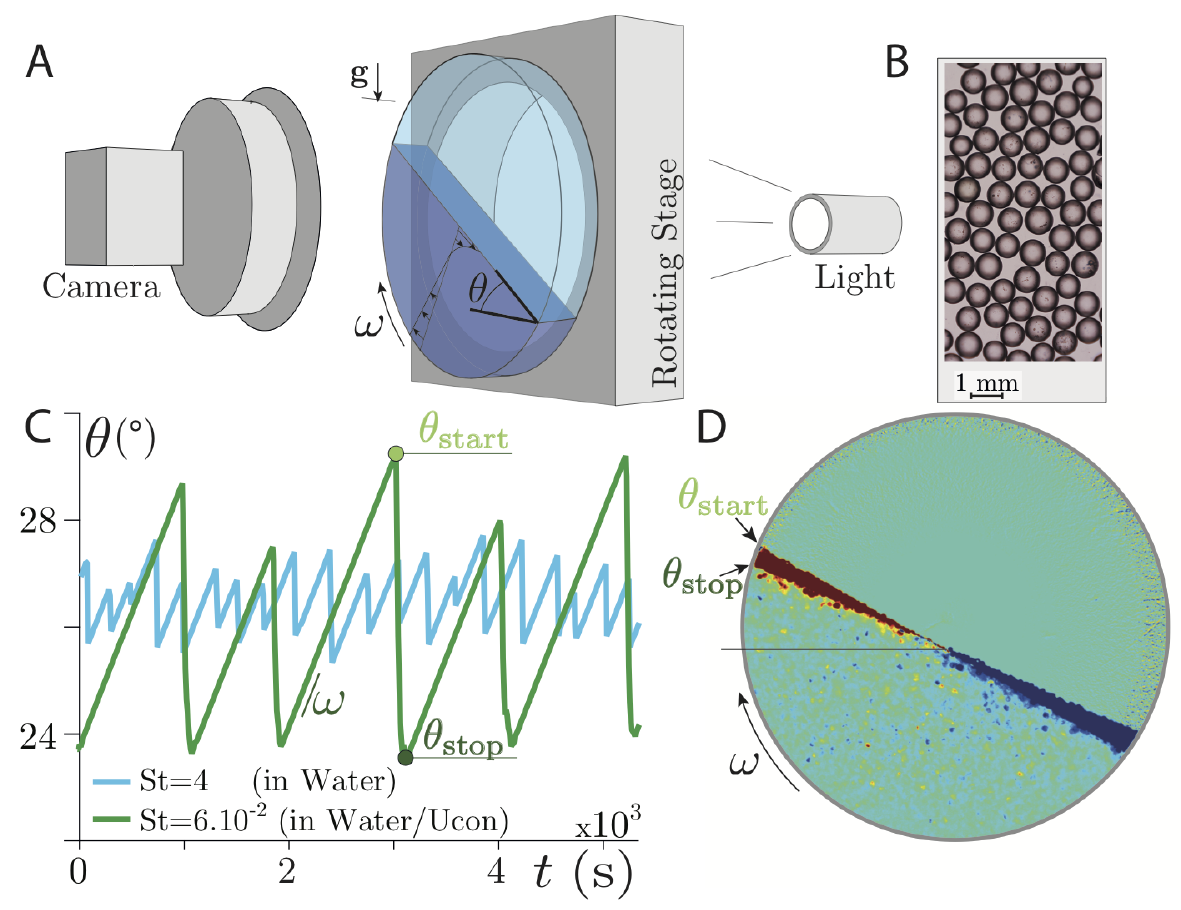}
\caption{Evidence of hysteresis in over-damped granular suspensions: (\textit{A}) Sketch of the experimental set-up. (\textit{B}) Picture of the large frictional glass beads ($d=490\pm 70\;\rm{\mu m}$, $\rho_p=2500\;$kg.m$^{-3}$). (\textit{C}) Angle of avalanche $\theta$ versus time for large glass beads immersed in pure water ($St= 4$) or in a mixture of water and Ucon oil ($St= 6\times 10^{-2}$); the rotation rate of the drum is $\omega = 5\times10^{-3}\;^{\circ}$.s$^{-1}$. (\textit{D}) Difference between two images taken just prior to and after an avalanche event.}
\label{Fig1}
\vspace{-1em}
\end{figure}

We first examine the behavior of an immersed granular pile of large glass particles ($d\approx 500\;\rm{\mu m}$) using the classic rotating drum configuration (see Fig. \ref{Fig1}\textit{A,B} and Appendix A: Materials and Methods). By imposing a slow and constant rotation rate $\omega$, the non-buoyant grains at the surface of the pile flow under their own weight forming an avalanche of angle $\theta$ on top of a region experiencing a rigid rotation with the drum. At low enough rotation rate, the avalanche dynamics is found to be unsteady, grains being either at rest or flowing. As shown in Fig. \ref{Fig1}\textit{C}, when the grains stand still, the pile angle $\theta$ increases at the rate $\omega$ set by the rotation rate of the drum. When the pile angle reaches $\theta_{\rm start}$, an avalanche is spontaneously triggered inducing a rapid downward surface flow of the grains: the pile angle decreases until it reaches $\theta_{\rm stop}<\theta_{\rm start}$. From there, subsequent cycles repeat on and on.  The resulting saw-tooth shape of the pile angle versus time is the phenomenological signature of hysteresis \cite{Courrech,forterre2008flows}. Its typical magnitude can be appreciated in Fig. \ref{Fig1}\textit{D} showing the difference between two images taken prior to and shortly after an avalanche. Note that this hysteresis does not arise from a specific preparation protocol or pre-compaction of the granular pile. 

The above phenomenology has been widely observed for inertial granular flows, both for aerial avalanches and for large grains immersed in low viscosity fluids \cite{allen1970,jaeger1989relaxation,rajchenbach1990,caponeri1995dynamics,Courrech}.  In agreement with \cite{Courrech}, we find that for grains immersed in water ($St= 4$), the hysteresis amplitude $\Delta \theta = \theta_{\rm start}-\theta_{\rm stop} \approx 1^\circ$ (see Appendix, Fig. \ref{Fig5} for comparison with Courrech du Pont et al data). However, in a much more viscous mixture of water and Ucon oil ($St= 6\times 10^{-2}$), the hysteresis amplitude $\Delta \theta \approx 4^\circ$ is significantly larger. This observation conflicts with the idea that hysteresis should vanish in fully overdamped suspensions \cite{Courrech}.   It also suggests that the Stokes number, which compares inertia to viscous effects, is not  the sole parameter that controls  the amplitude of hysteretic avalanches  in immersed granular media.

\subsection{Probing hysteresis by tuning microscopic friction}

\begin{figure*}
\centering \includegraphics[width=0.75\linewidth]{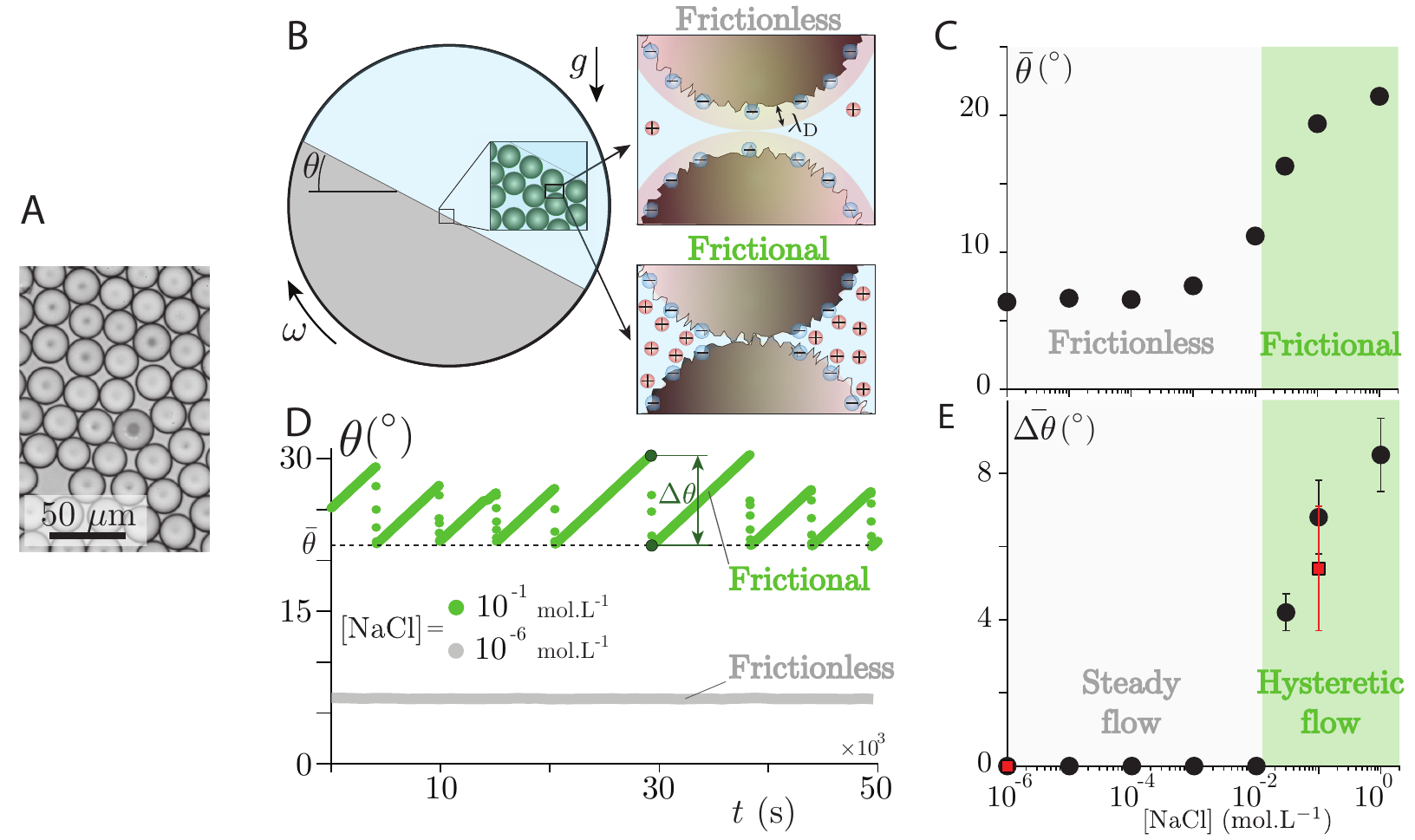}
\caption{Interparticle friction triggers hysteretic avalanches: (\textit{A}) Picture of the silica particles ($d=23.46 \pm 1.06\;\rm{\mu m}$, $\rho_p=1850\;$kg.m$^{-3}$, $St =2 \times10^{-2}$). (\textit{B}) Schematic depicting how interparticle friction is tuned by varying the solvent ionic concentration. (\textit{C}) Time averaged avalanche angle $\bar{\theta}$ versus ionic concentration $[\rm{NaCl}]$. (\textit{D}) Avalanche angle $\theta$ versus time for frictionless (black) and frictional (green) particles. (\textit{E}) Mean hysteresis amplitude $\bar{\Delta \theta}$ (averaged over 15 avalanches) versus salt concentrations. Square markers correspond to experiments performed with a water/glycerol mixture ($St \!= \! 2\times10^{-3}$). All measurements were performed for $\omega \!=\!10^{-3}\;^{\circ}$.s$^{-1}$ except for the water/glycerol mixture where $\omega\!=\!10^{-4}\;^{\circ}$.s$^{-1}$. }
\label{Fig2}
\vspace{-1em}
\end{figure*}

Besides inertia, a key ingredient put forward in the literature to account for the emergence of hysteresis   is the interparticle friction \cite{PeyneauRoux2008,DeGiuliPNAS}. To investigate the effect of this parameter, we use a suspension composed of non-Brownian silica particles ($d\approx 24\;\rm{ \mu m}$) (Fig. \ref{Fig2}\textit{A}) which, as we have recently shown \cite{Clavaud2017}, allows the control of the interparticle friction forces (Appendix A: Materials and Methods). When immersed in water, silica particles spontaneously develop negative surface charges, which generate an electrostatic repulsive force between the grains \cite{Israelachvili2011Book}.  This force, under low confining stress (here set by the weight of the flowing granular layer), prevents the particles from making solid contact because the range of the repulsive force, i.e., the Debye length $\lambda_D$, is larger than the particle roughness (Fig. \ref{Fig2}\textit{B} top); in this case the particles behave as if they were frictionless. Conversely, interparticle friction can be turned on simply by dissolving electrolytes (NaCl here) in water, which screens the surface charges. This decreases $\lambda_D$ which eventually becomes smaller than the particle roughness (Fig. \ref{Fig2}\textit{B} bottom); solid frictional contact between particles is thereby activated above a critical ionic concentration. 

This transition between frictionless and frictional particles is illustrated  in Fig. \ref{Fig2}\textit{C} showing the evolution of the quasi-static mean pile avalanche angle $\bar\theta$ as a function of the ionic concentration. For low salt concentration, [NaCl]$< 10^{-3}$ mol.L$^{-1}$, the mean avalanche angle is very small, $\bar \theta\approx 6^{\circ}$, a value remarkably close to that obtained numerically for ideal frictionless spheres, $\theta= 5.76^{\circ}$ \cite{PeyneauRoux2008}. Conversely, when further increasing the salt concentration, particles start making solid contact therefore involving friction in the avalanche dynamics: the avalanche angle increases to reach $\bar \theta\approx 22^{\circ}$ at large salt concentrations, a value typical for frictional grains \cite{Courrech,forterre2008flows}. 

The most important finding here is that the interparticle friction not only affects the mean avalanche angle but also the hysteretical nature of the flow (Fig. \ref{Fig2}\textit{D}). When particles interact through frictional contacts (large ionic concentration [NaCl]$=10^{-1}$ mol.L$^{-1}$), one observes large hysteretic avalanches (with a saw-tooth shape) similar to those obtained previously with the macroscopic beads immersed in a viscous mixture (Fig. \ref{Fig1}\textit{C}). By contrast, when interparticle friction is turned off (low ionic concentration [NaCl]$=10^{-6}$ mol.L$^{-1}$), hysteresis completely disappears. Grains at the surface of the pile then flow steadily at a fixed avalanche angle. The measure of the amplitude of the hysteresis $\Delta \theta=\theta_{\rm start}-\theta_{\rm stop}$ for various ionic concentrations confirms that hysteresis relies on the presence of interparticle friction (Fig. \ref{Fig2}\textit{E}). No hysteresis is observed as long as interparticle friction is turned off ([NaCl]$< 10^{-3}$ mol.L$^{-1}$), while large hysteresis appears when the suspension becomes frictional ([NaCl]$> 10^{-2}$ mol.L$^{-1}$), with an amplitude that increases with the salt concentration. Importantly, the Stokes number for all these experiments is very small ($St = 2\times10^{-2}$) and identical.  Yet, the frictional suspensions exhibit hysteresis while the frictional suspensions do not. This result confirms that inertia is not necessary to observe hysteresis  in granular suspensions, while interparticle friction is (see also Appendix, Fig. \ref{Fig5}).

It is worth noting that in our system the emergence of hysteresis cannot be attributed to adhesive forces between particles. Firstly, for all ionic concentrations, avalanche angles have a constant slope from the top to the bottom of the avalanche, unlike adhesive powders. Secondly, supplementary experiments where performed in a $36/64\%$ water/glycerol mixture in order to match, down to the second digit, the index of refraction of the suspending fluid to that of the silica beads. With such an index matching, adhesive van der Waals forces are lowered by about two orders of magnitude compared to those expected in water \cite{Israelachvili2011Book}. Yet, the results remain similar: no hysteresis is observed at low ionic concentrations and similar hysteresis amplitudes are found at large ionic concentrations (square markers in Fig. \ref{Fig2}\textit{E}). 

\subsection{Link between hysteresis and rheology}

\begin{figure*}[t!]
\centering \includegraphics[width=0.95\textwidth]{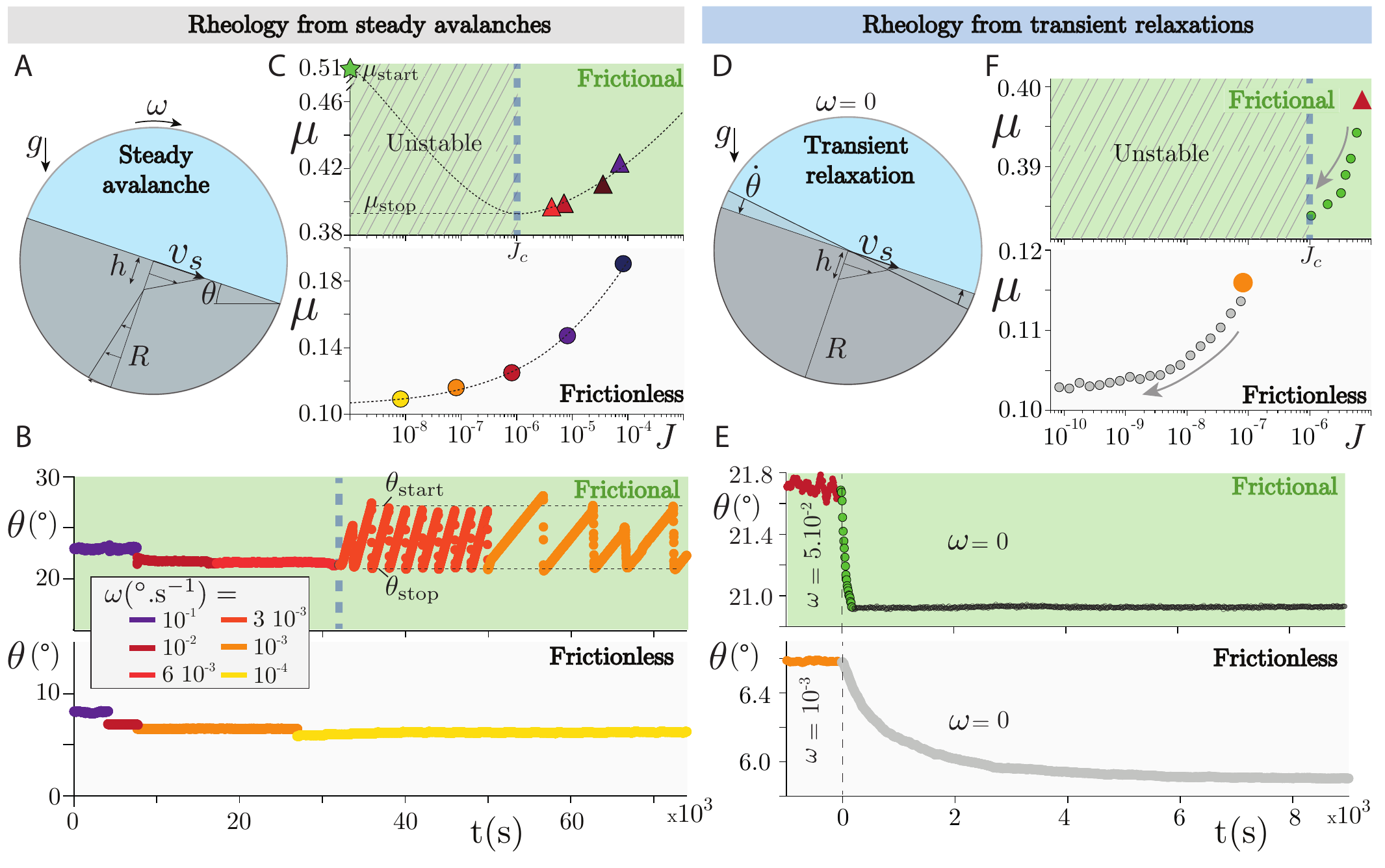}
\caption{Frictional and frictionless rheologies extracted from steady avalanches (\textit{A--C}) and transient relaxations (\textit{D--F}). (\textit{A}) Sketch defining the parameters used to extract the rheology for steady avalanches.  (\textit{B}) Avalanche angle $\theta$ versus time for frictional and frictionless silica particles ([NaCl]$=10^{-1}$ and $10^{-6}$ mol.L$^{-1}$, respectively) measured by successively decreasing the rotation rate $\omega$ of the drum. (\textit{C}) Corresponding suspension effective friction coefficient $\mu$ versus viscous number $J$; the color code corresponds to the rotation rate of the drum. (\textit{D}) Sketch defining the parameters used to extract the rheology from transient relaxations. (\textit{E}) Relaxation of the angle of avalanche versus time for frictional and frictionless silica particles ([NaCl]$=10^{-1}$ and $10^{-6}$ mol.L$^{-1}$, respectively). The preparation rotation rates are $\omega=5\times10^{-2}\;^{\circ}$.s$^{-1}$ and $\omega=10^{-2}\;^{\circ}$.s$^{-1}$ respectively, the drum is stopped ($\omega=0$) at $t=0$. (\textit{F}) Corresponding $\mu$ versus $J$. To compute $J=  \eta |\dot{\theta} | R^2 / (h^2 P)$ from $\theta(t)$ measured in \textit{E}, data are smoothed using time-windows of variable widths and  $\dot\theta(t)$ is computed using finite-difference.  The same color code as in \textit{E} is used.}
\label{rheol}
\vspace{-1em}
\end{figure*}
In solid friction and geophysics flows, hysteresis and stick-slip dynamics are often related to a velocity-weakening regime of the flow near the onset of motion, i.e., a shear stress that decreases with the deformation rate. To understand the emergence of hysteresis in overdamped suspensions, we further analyze the dynamics of the avalanches with the aim of extracting the effective rheology of both the frictional and frictionless silica beads (immersed in solutions of ionic concentration [NaCl]$=10^{-1}$ and $10^{-6}$ mol.L$^{-1}$, respectively).

Under steady flow conditions, the effective rheology of the suspension can be obtained as follows.  Momentum balance at the free surface implies that the macroscopic friction coefficient of the suspension $\mu$ (the ratio of the tangential to normal stresses) is directly given by the avalanche angle $\theta$:  $\mu=\tan \theta$ \cite{jop2005crucial}.  Moreover, as illustrated in Fig. \ref{rheol}\textit{A}, conservation of mass implies that the downward flux of grains during steady avalanching is balanced by the upward motion of grains experiencing rigid rotation with the drum, yielding  $h v_s \sim \omega R^2$, where $h$ is the thickness of the flowing layer, $v_s$ the surface velocity of the avalanche and $R$ the radius of the drum. The typical shear rate within the flowing granular layer is $\dot{\gamma}\sim v_s/h$. The effective viscous number $J=\eta \dot{\gamma}/P$ characterizing the flow \cite{boyer2011unifying} can therefore be computed as $J =  \eta \omega R^2 / (h^2 P)$ with $P=\phi \Delta \rho g h \cos \theta$ and the volume fraction $\phi \approx 0.6$. In these expressions, the sole unknown is the thickness of the flowing layer $h$. In the quasi-static regime of interest here, $h$ is mainly set by the grain diameter \cite{GDRmidi2004}. For simplicity, we use $h=50d$ for frictionless particles and  $h=25d$ for frictional particles, as suggested by our experiments (Appendix, Fig. \ref{Fig6}).

Measuring the steady avalanche angle $\theta$ and the rotation rate $\omega$ of the drum therefore gives access to the effective frictional rheology $\mu(J)$ of the suspensions\footnote{Note that the friction coefficient $\mu$ is measured at the free surface while the viscous number $J$ corresponds to an average over the thickness of the flowing layer.}. Fig. \ref{rheol}\textit{B} shows the steady avalanche angle obtained for various rotation rates for both frictional and frictionless particles. At large rotation rates both systems exhibit steady flows with stationary angles that progressively decrease as the rotation rate is decreased. These data are used to extract the rheologies $\mu(J)$ in  Fig. \ref{rheol}\textit{C} for the frictional (top) and frictionless (bottom) grains. Each point corresponds to a different drum rotation rate (same color code as in Fig. \ref{rheol}\textit{B}). An important difference between these two systems is that below the critical rotation rate $\omega_c=6\times10^{-3}\;^{\circ}$.s$^{-1}$, the flow of frictional grains becomes hysteretic while frictionless grains flow steadily down to the smallest rotation rate accessible experimentally. This reflects a strong difference of rheologies. Indeed, for the frictionless suspension, as the rotation rate of the drum is decreased, the effective macroscopic friction coefficient of the suspension $\mu(J)$ decreases monotonously until reaching $\mu_0\approx 0.1$ as $J\rightarrow 0$, Fig. \ref{rheol}\textit{C} (bottom). This rheological flow law belongs to those of velocity-strengthening materials where friction grows with the deformation rate; this type of rheology yields stable flows. The behavior of frictional particles is markedly different: below $\omega_c$ corresponding to a finite $J_c= \eta \omega_c R^2 / (h^2 P)$, the flow becomes unstable and hysteretic avalanches come into play. In this regime, once grains are at rest, starting the flow requires the pile angle to reach  $\theta_{\rm start}$. This angle defines a starting friction coefficient $\mu_{\rm start}\approx 0.51$ much larger than $\mu_{\rm stop}=\mu(J_c)\approx 0.39$.  These findings suggest that the rheology $\mu(J)$ of frictional particles is a non-monotonous function of $J$ with a velocity-weakening region below $J_c$, as illustrated by the dashed line in Fig. \ref{rheol}\textit{C} (top). Such multivalued rheological flow law is known to give rise to instabilities under loading, including hysteretic avalanches \cite{peyneauPhD}.

Steady avalanches measurements only provide the rheology for discrete values of $J$ (corresponding to the imposed rotation rates $\omega$). A way to obtain continuous measurements and further refine the rheology $\mu(J)$ close to the flow onset, is to analyze the transient relaxations. To do so, the system is first prepared in a steady regime at a given rotation rate $\omega$ (larger than $\omega_c$ for frictional particles). The drum is then suddenly stopped ($\omega=0$), and the relaxation of the pile angle $\theta(t)$ is recorded until the  flow stops (Fig. \ref{rheol}\textit{E}). Assuming these relaxations are quasi-stationary, the macroscopic friction coefficient of the suspension $\mu$ is again given by the avalanche angle $\theta$ using the relation $\mu=\tan \theta$. Moreover, mass conservation, now that the drum is stopped ($\omega=0$), implies that the downward grains flux $h v_s$, leading to the pile angle variation $\dot{\theta}=d\theta / dt$, obeys $h v_s\sim  -R^2 \dot{\theta} $ (Fig. \ref{rheol}\textit{D}). The viscous number is then obtained as $J=  \eta | \dot{\theta}|  R^2 / (h^2 P)$. 

The relaxations of frictional (Fig. \ref{rheol}\textit{E} top) and frictionless (Fig. \ref{rheol}\textit{E} bottom) particles are found to strongly differ both in timescales and shapes. For frictionless grains, the angle of the pile relaxes slowly, asymptotically reaching its final value after about 8000 s. Conversely, for the frictional grains, the pile angle relaxes in a much shorter time (about 50 s) and its final value is reached with an abrupt change in the relaxation dynamics:  the relaxation rate of the pile angle sharply transitions from a finite value to zero, yielding a sudden stop of the flow.  Again, these markedly different features for the relaxations are the outcome of the intrinsically different rheological laws shown in Fig. \ref{rheol}\textit{F}. For frictionless grains, the rheology remains monotonous two decades bellow the lowest viscous number investigated with the steady-state measurements. Conversely, the rheology obtained for frictional grains confirms the existence of a critical viscous number $J_c$ below which no flow is possible. As the pile angle progressively relaxes, the transient avalanche stops abruptly when the system reaches the viscous number $J=J_c$.  In Appendix B, we show that the discontinuity of $\dot{\theta}$ observed in Fig. \ref{rheol}\textit{D} (top) is inherently a consequence of the finite value of $J_c$. 

\begin{figure}[t!]
\centering \includegraphics[width=0.95\linewidth]{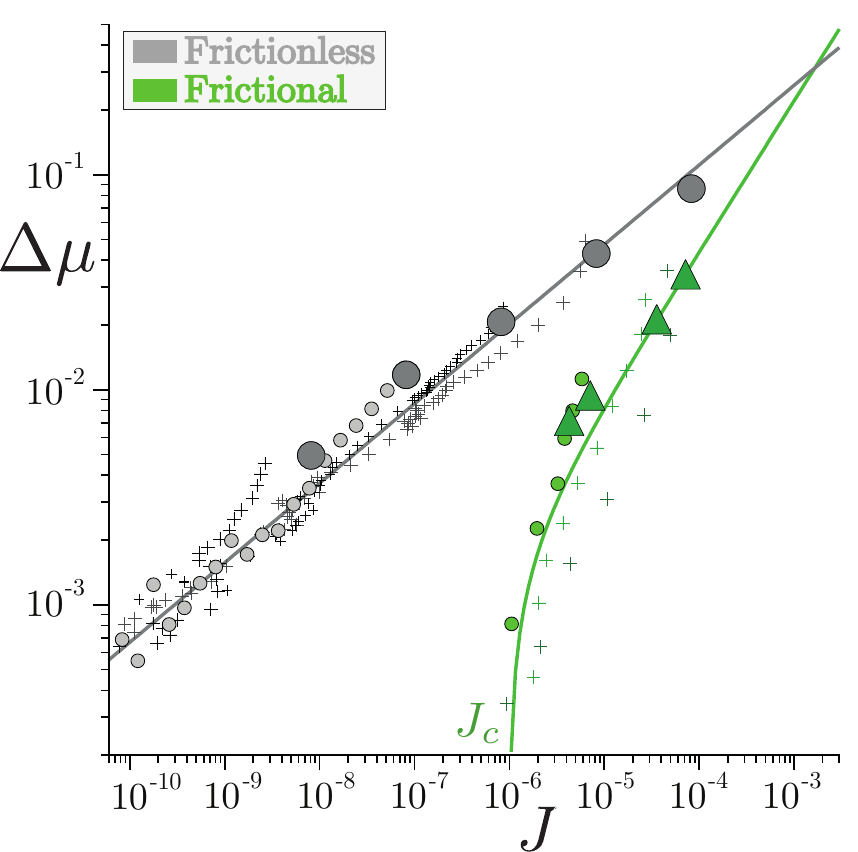}
\caption{Reduced macroscopic friction coefficient $\Delta \mu = \mu - \mu_0$ versus viscous number $J$ for frictional and frictionless particles immersed in solutions of ionic concentration ([NaCl]$=10^{-1}$ and $10^{-6}$ mol.L$^{-1}$, respectively). Large markers were obtained from the steady-state measurements presented in Fig. \ref{rheol}\textit{A--C}. Small markers correspond to transient relaxations measurements from different initial steady states as described in Fig. \ref{rheol}\textit{D--F}. Solid lines are fits  (in grey) by $\Delta \mu=(J/J_0)^{\alpha}$ with $\alpha=0.37 \pm 0.05$ and $J_0\approx2.27\times 10^{-2}$ for frictionless particles and (in green) by $\Delta \mu=((J-J_c)/J_0)^{\beta}$ with $\beta= 0.7 \pm 0.3$, $J_0 \approx8.69\times10^{-3}$ and $J_c \approx 10^{-6}$ for frictional particles.} 
\label{final}
\end{figure}

To conclude, Fig. \ref{final} gathers the rheological laws obtained both from analyzing steady state avalanches and transient relaxations. The reduced macroscopic friction coefficient $\Delta \mu=\mu-\mu_0$ is plotted versus $J$ where $\mu_0=\mu(J\rightarrow0) \approx 0.1$ for frictionless particles and $\mu_0=\mu_{\rm stop}\approx 0.39$ for frictional particles\footnote{The asymptotic values $\mu_0$ and $\mu_{\rm stop}$ weakly depend on the rotation rate of preparation before the drum is stopped. Exact values are given in Appendix, Fig. \ref{Fig7}.}. The large symbols correspond to the steady-state measurements. We find that all data collapse on two separate master curves distinguishing frictionless and frictional grains. The data for frictionless grains follows the monotonous constitutive relation $\Delta \mu \sim  J^{\alpha}$ with $\alpha=0.37 \pm 0.05$ over six decades of viscous number $J$. The value of the exponent is remarkably close to previous numerical observations \cite{peyneauPhD,olsson2011critical,lerner2012,degiuli2015unified}  as well as  the theoretical prediction 0.35 \cite{degiuli2015unified}. To our knowledge, this is the first experimental validation of these predictions. By contrast, the rheology of frictional grains exhibits a minimum at $J=J_c$ and, for $J > J_c$, can be well fitted with $\Delta \mu \sim  (J-J_c)^{\beta}$ with $\beta= 0.7 \pm 0.3$ and $J_c \approx 10^{-6}$.  

\section{Discussion}

The hysteresis observed at the onset of granular flows has generally been explained by invoking mechanisms based on inertia, such as dissipation by shocks \cite{quartier2000dynamics,Courrech}, or endogenous acoustic noise arising from collisions \cite{DeGiuliPNAS}. In this article, by studying avalanches of viscous suspensions in rotating drums, we show that large hysteresis also occurs in overdamped systems where inertial effects are completely negligible. We show however that the presence of interparticle friction is crucial  to observe hysteresis.  Avalanches of frictional particles become hysteretic below a critical rotation rate of the drum, while for frictionless particles steady flow is always observed. Examination of the avalanche dynamics, both in steady state and during transient relaxations, reveals the difference between the rheological laws of frictionless and frictional suspensions. Frictionless particles exhibit a monotonic behavior while the law for frictional particles is non-monotonic.

These rheological laws can be used to rationalize, within a simple model, the main features of the avalanches observed for both types of grains (frictionless and frictional) (see Appendix).  \textit{(i)} When the drum is rotated continuously, steady flows are possible only if the flow rule is stable, i.e., velocity-strengthening. This is satisfied for all values of $J$ (or corresponding $\omega$) for frictionless grains but only above $J_c$  (or corresponding $\omega_c$) for frictional particles.  Below this critical point hysteresis emerges (Fig. \ref{rheol}\textit{B} and Appendix, Fig. \ref{Fig8}\textit{B}). \textit{(ii)} As the drum is stopped, the avalanche angle relaxes asymptotically to its final value for frictionless particles because the flow law is monotonic. By contrast, the avalanche angle for frictional particles relaxes in a finite time according to their non-monotonic flow law (Fig. \ref{rheol}\textit{E} and Appendix, Fig. \ref{Fig8}\textit{C}). \textit{(iii)}  The rheology of frictionless grains follows a power law with a lowest exponent than that obtained for frictional particles ($\alpha<\beta$). This has a somewhat counter-intuitive consequence since it implies that, at low viscous number, for the same friction difference $\Delta \mu$, frictionless grains must flow more slowly (smaller $J$) than frictional grains. This explains the long and short relaxation times observed for frictionless and frictional grains respectively (Fig. \ref{rheol}\textit{E} and Appendix, Fig. \ref{Fig8}\textit{D}).

Our interpretation of the transition between discontinuous and continuous flow in rotating drums differs from early interpretations based on a balance between the rotation time scale and the avalanche duration  \cite{rajchenbach1990} or on the stochastic fluctuation of the friction law \cite{fischer2009transition}. Here the critical rotation rate $\omega_c$ reflects the critical viscous number $J_c$ below which the friction law is velocity-weakening and therefore unstable. Note that the rotating drum, which is a stress-driven configuration, does not give access to the unstable branch of the rheology. In dry granular materials, few direct evidences of a velocity-weakening regime exist \cite{dijksman2011jamming,kuwano2013crossover,DeGiuliPNAS} and other works have suggested the existence of a minimum in the friction law  \cite{caponeri1995dynamics,pouliquen2002friction,edwards2017formation}. In dense suspensions, measurements and numerical simulations of the $\mu(J)$ rheology have so far reported only monotonic laws \cite{boyer2011unifying,trulsson2012transition,amarsid2017viscoinertial,trulsson2017effect}. However,  very few data are available for viscous numbers smaller than the critical viscous number reported here, $J_c \approx 10^{-6}$. An important question is to understand the physical parameters that set this value of $J_c$.  One possibility could be that $J_c$ is size-dependent and vanishes for large systems, as was suggested in the case of inertial grains \cite{pouliquen2002friction}. Another possibility could be that $J_c$ is related to an intrinsic timescale not included in the hard sphere limit. For instance, the collision timescale introduced in the endogenous noise mechanism for inertial particles \cite{DeGiuliPNAS} could be replaced by a viscoelastic rearrangement time for overdamped systems. 

Whatever the mechanism, our work shows that solid contact and interparticle friction is required to observe hysteresis in dense suspensions, as in dry materials \cite{PeyneauRoux2008,DeGiuliPNAS}. Interestingly, the dilatancy exhibited by granular materials at the onset of the flow (Reynold dilatancy) is also a signature of friction between particles \cite{PeyneauRoux2008,Clavaud2017}. Thus, dilatancy and hysteresis could share the same origin, an idea already found in  Bagnold's work \cite{bagnold1966shearing}.  On the other hand, solid friction is also known to involve aging phenomena and velocity-weakening behaviors at low sliding velocities \cite{baumberger2006solid}. It is thus likely that the hysteresis observed at a macroscopic level is strongly affected by the hysteresis of solid friction at the contact level.   

Overall, our study reveals that the flow onset in dense suspensions is hysteretical and that this hysteresis stems from the presence of interparticle friction. This finding provides another strong evidence that solid contacts are crucial to understanding the dynamics of suspensions close to the jamming transition, corroborating recent advances in the field \cite{guazzelli2018rheology}. In the geophysical context, our results may help to understand the failure mechanism  of undersea landslides and better predict the occurrence of massive events \cite{lovholt2017}. Interestingly, both over-damped suspensions and inertial granular materials are found to exhibit similar hysteretical signatures.  Whether this hysteresis arises from collective effects  or simply reflects the intrinsic hysteresis of solid friction at the scale of particle contacts is an  important open question. To address this issue, it would be interesting to extend current discrete simulations of dense suspensions \cite{amarsid2017viscoinertial,trulsson2017effect} to account for hysteresis at the scale of particle contact.

\begin{acknowledgements}
We thank O. Pouliquen and M. Trulsson for discussions, P. Dame and E. Fernandez for performing preliminary experiments, S. No\"el and F. Ratouchniack for help in building the experimental set-up. M. W. and H. P. thanks the Swiss National Science Foundation for support under Grant No. 200021-165509 and the Simons Foundation Grant (No. 454953 Matthieu Wyart). This work was supported by the European Research Council under the European Union Horizon 2020 Research and Innovation programme (ERC grant agreement No. 647384 ), by the Labex MEC (ANR-10-LABX-0092) under the A*MIDEX project (ANR-11-IDEX-0001-02) funded by the French government program Investissements d'Avenir.
\end{acknowledgements}

\appendix

\section{Materials and Methods}

\subsection*{Particles}
The grains used in Fig. 1 are large glass beads of diameter $d=490\pm70\;\rm{\mu m}$ and density $\rho_p =2500\;\rm{kg.m^{-3}}$. The silica beads used in Figs. 2--4 are commercial particles from Microparticles GmbH with diameter $d=23.46 \pm 1.06\;\rm{\mu m}$ and density $\rho = 1850\;\rm{kg.m^{-3}}$.

\subsection*{Rotating Drum Experiments}
The drum used in Fig. 1 (for the large glass beads) has a diameter of 52 mm and a depth of 10 mm with a coarsened side wall and is made out of plexiglass plates. It is filled either with pure (microfiltered) water or with a mixture of Ucon oil and water of viscosity $\eta_f = 57$ mPa.s and density $\rho_f = 1005\;$kg.m$^{-3}$. The drum used in Fig. 2--4 (for the silica particles) has a diameter of 12 mm and a depth of 3 mm with a coarsened side wall. The front and back walls are both made out of silica slides. The drum is filled with various ionic (NaCl) aqueous solutions. Both drums are half filled with particles and fully filled with the liquid. The drums are mounted on a precision rotating stage (M-061PD from PI piezo-nano positioning). Images are acquired with a 2048x1080 Basler digital camera with a resolution of $\sim10\;\rm{\mu m/pix}$. The angle of avalanche is measured with a precision of $0.01^\circ$ using a sub-pixel detection of the interface between grains and liquid.

\subsection*{Cleaning and preparation protocol}
Prior to performing experiments, both the silica particles and the silica walls of the drum are first  cleaned in Piranha solution (1:2 of $\rm{H_2O_2}$:$\rm{H_2SO_4}$) for 10 minutes, then rinsed several times with pure micro-filtered water. They are then immersed in the desired ionic solution, placed in a ultrasonic bath for 10 minutes and rinsed 4 times with the ionic solution. After testing that the suspending fluid conductivity corresponds to the desired ionic concentration, the grains are immediately placed in the drum. Note that the amplitude of hysteresis $\Delta \theta$ can depend significantly on the cleaning and preparation protocol. However, the low ionic concentration water cases are less sensitive (hysteretic avalanches were never observed for [NaCl]$< 10^{-3}$ mol.L$^{-1}$).

\section{Average amplitude of hysteresis $\bar{\Delta \theta}$ versus Stokes number $St$}
\begin{figure}[h!]
\centering \includegraphics[width=1.0\linewidth]{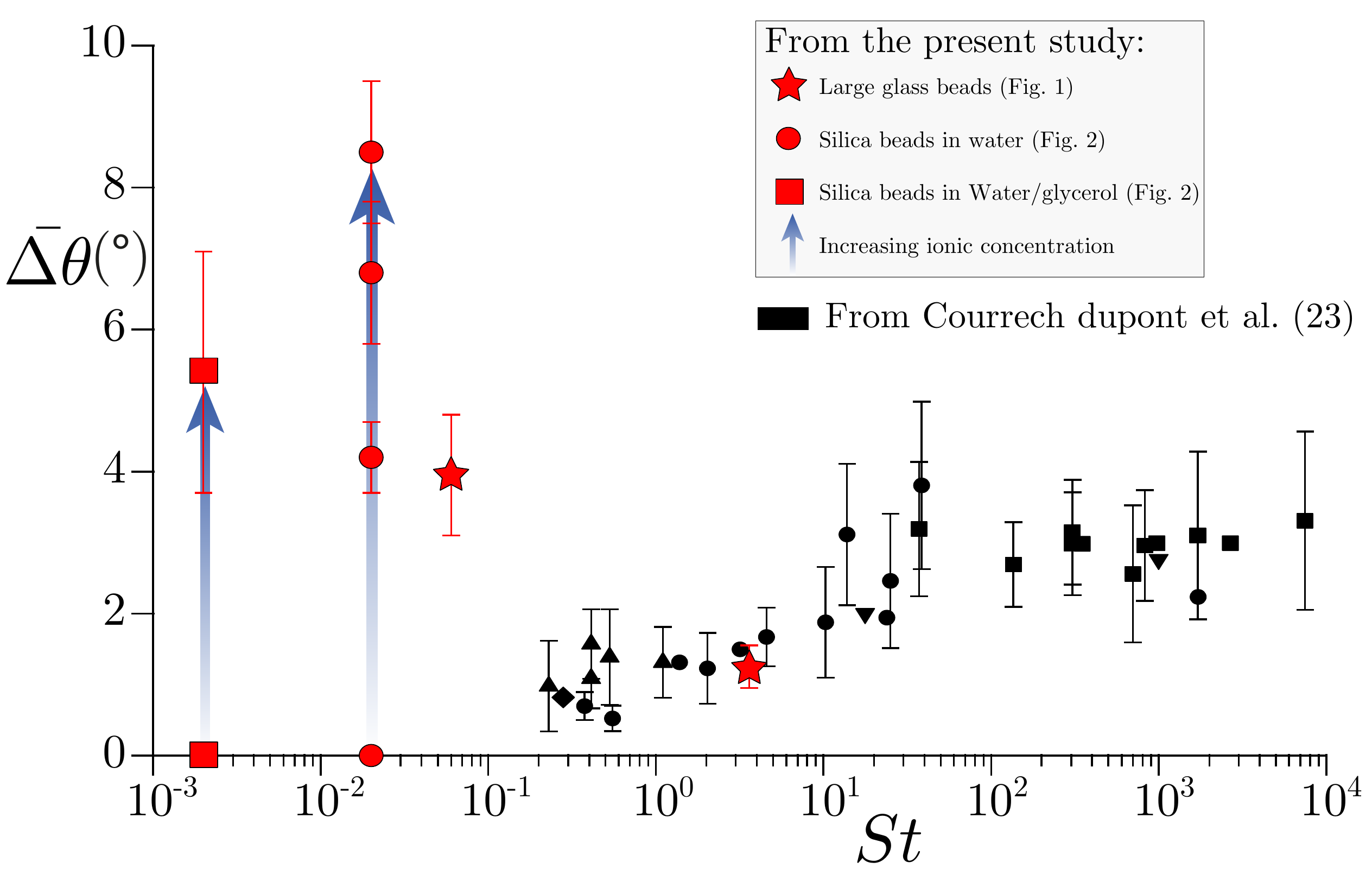}
\caption{Average amplitude of hysteresis $\bar{\Delta \theta}$ versus Stokes number $St$. Comparison between data obtained by Courrech du Pont et al (23) (in black) and in the present study. Red star markers correspond to the large glass beads in water and in the mixture of Ucon-oil  and water (Fig. 1\textit{C}). Red circle (resp. square) markers correspond to the silica particles in different salt concentration in water (resp. water/glycerol) (Fig. 2\textit{E}). For the present study hysteresis amplitudes were measured at a rotation rate of $\omega\!=\!10^{-3}\;^{\circ}$.s$^{-1}$, except for the silica particles in the water/glycerol mixture where $\omega\!=\!10^{-4}\;^{\circ}$.s$^{-1}$ and for the large glass beads where  $\omega\!=\!5\times10^{-3}\;^{\circ}$.s$^{-1}$.}
\label{Fig5}
\end{figure}

\section{Complementary measurements}
\subsection*{Flowing thickness measurements}
\begin{figure}[h!]
\centering \includegraphics[width=0.9\linewidth]{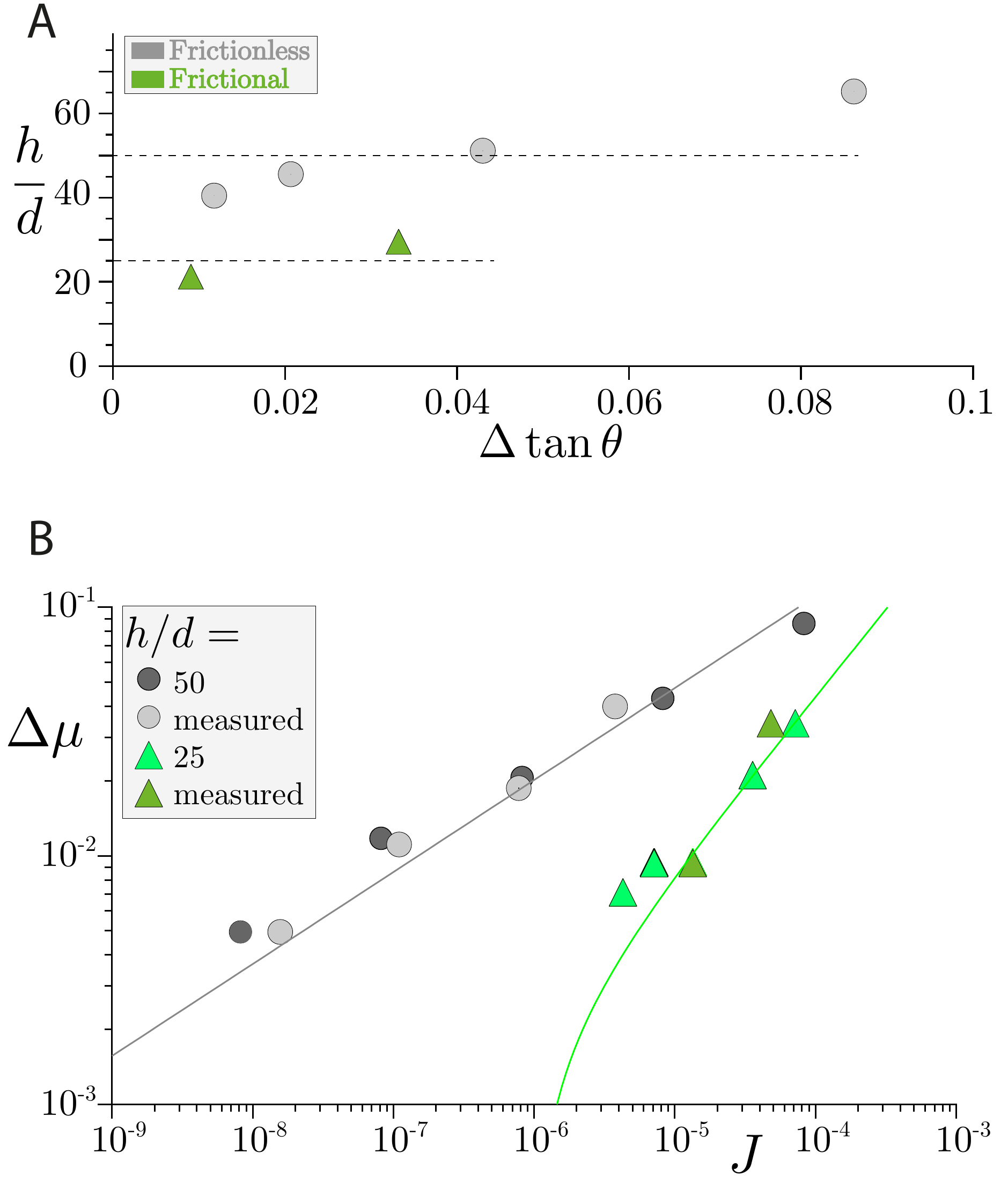}
\caption{(\textit{A}) Flowing thickness $h/d$ of the avalanche in the steady states measured at the front wall of the drum using Particle Image Velocimetry, as function of the reduced avalanche angle $\Delta \tan \theta$ for frictionless (grey circles,  [NaCl]$=10^{-6}$ mol.L$^{-1}$) and frictional (green triangles, [NaCl]$=10^{-1}$ mol.L$^{-1}$) particles (silica spheres of diameter  $d=23.46 \pm 1.06\;\rm{\mu m}$ and density $\rho_p=1850\;$kg.m$^{-3}$). Dashed lines correspond to average values $h/d= 25$ (frictional grains) and $h/d=50$ (frictionless grains). (\textit{B}) Rheology $\mu(J)$ extracted from the stationary regimes using the averaged value of $h/d$ or the measured values shown in Fig. \ref{Fig2}\textit{A} for computing the viscous number $J =  \eta \omega R^2 / (h^2 P)$ (grey symbols: frictionless grains, green symbols: frictional grains, same experimental conditions as in Fig. \ref{Fig2}\textit{A}).  }
\label{Fig6}
\end{figure}

In this section we provide complementary measurements to validate our approximations of constant flowing thickness. Fig \ref{Fig6}.\textit{A} reports the measured flowing thickness for both frictional and frictionless silica particles in the steady states. Fig \ref{Fig6}.\textit{B} compares the difference of rheological curves between the constant flowing thickness approximation and the consideration of the measured flowing thickness.

\subsection*{Asymptotic friction coefficients}
\begin{figure}[h!]
\centering \includegraphics[width=0.9\linewidth]{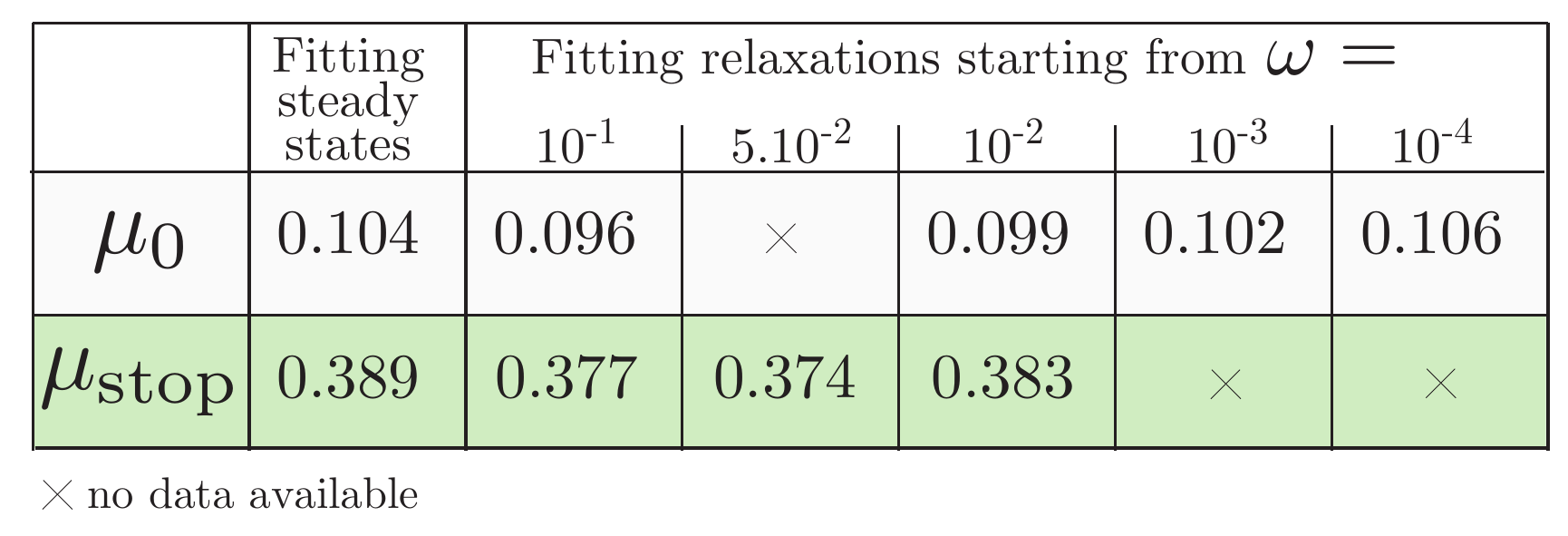}
\vspace{-0.5cm}
\caption{Values of $\tan \theta$ when the flow stops for frictionless (top, [NaCl]$=10^{-6}$ mol.L$^{-1}$) and frictional (bottom, [NaCl]$=10^{-1}$  mol.L$^{-1}$) particles (silica spheres of diameter  $d=23.46 \pm 1.06\;\rm{\mu m}$ and density $\rho_p=1850\;$kg.m$^{-3}$). The first column is obtained by fitting the data in steady rotations using a power law with a threshold. The other columns correspond to the angles measured at the end of the relaxations for different initial steady rotation rates.}
\label{Fig7}
\end{figure}
\section{Simple avalanche model in the rotating drum}

\begin{figure*}[t!]
\centering \includegraphics[width=0.95\textwidth]{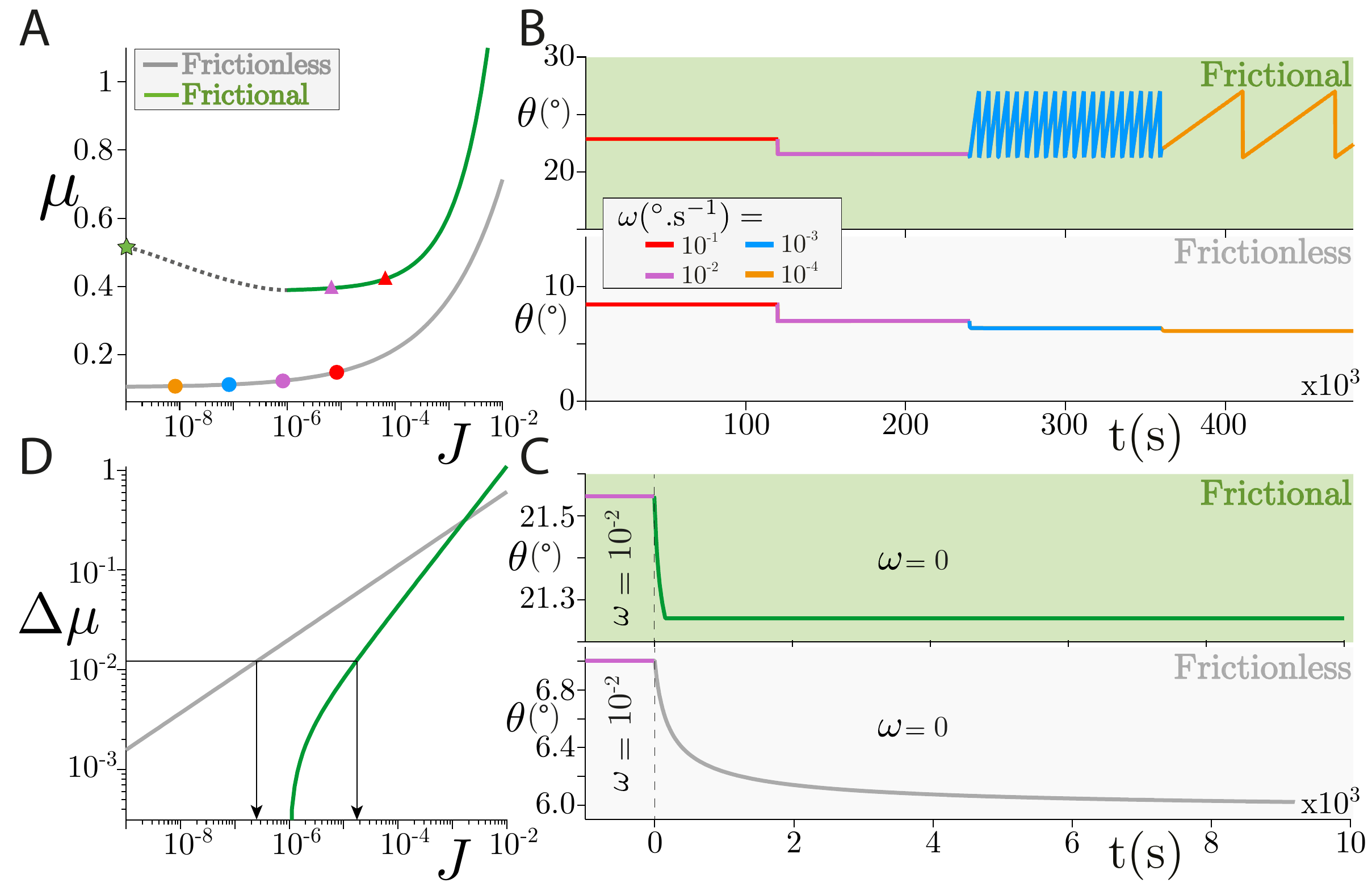}
\caption{Prediction of the avalanche model in the rotating drum. (\textit{A}) Rheologies for the frictionless (grey curve, Eq. [\ref{eqrheonofric}]) and frictional (green curve, Eq. [\ref{eqrheofric1}]) suspensions as deduced from the experiments. The symbols correspond to the steady states for the constant rotation rates given in \textit{B}. (\textit{B}) Time evolution of the avalanche angle for decreasing steps of rotation rates. (\textit{C}) Relaxation of the avalanche angle when the drums stops ($\omega=0$) at $t=0$ after a continuous rotation. (\textit{D}) Reduced macroscopic friction coefficient $\Delta \mu = \mu - \mu_0$ versus viscous number $J$ showing that for a given $\Delta \mu$, the viscous number for the frictional suspension is higher than for the frictionless suspension, thereby explaining the faster relaxations observed for the frictional case.}
\label{Fig8}
\vspace{-1em}
\end{figure*}

Here we derive a simple model to predict the avalanche dynamics in the rotating drum for the frictional and frictionless suspensions. For simplicity, the surface flow velocity $v_s$ and thickness $h$ of the avalanche is assumed uniform. In this case, mass conservation implies that the temporal variation of the angle of the pile of grains $\theta(t)$ is increased by the upward motion of static grains due to the solid rotation of the drum and decreased by the downward motion of grains due to the flow, that is
\begin{eqnarray}
\frac{d\theta}{dt} \approx - \frac{h v_s}{R^2} + \omega
\label{eq1}
\end{eqnarray}
where $R$ is the radius of the drum and $\omega$ its rotation rate. Within the quasi-stationary approximation, the momentum balance at the free surface of the pile implies that  $\mu=\tan \theta$, where the $\mu$ is the macroscopic friction coefficient of the suspension (ratio of shear to normal stress). Finally, the constitutive relation of the suspension is $\mu=\mu(J)$, where  $J=\eta \dot \gamma / P$ is the viscous number, $\eta$ is the  fluid viscosity, $\dot \gamma\approx v_s/h$ is the shear rate and $P\approx \phi \Delta \rho g h \cos\theta$ is the pressure, with $\phi\approx 0.6$ the packing fraction, $\Delta \rho$ the particle density minus the fluid density and $g$ the gravity. The time evolution of the avalanche angle is then given by: 
\begin{eqnarray}\label{eqdiff}
\frac{d\theta}{dt} \approx - \frac{h^3 \Delta\rho g \phi \cos\theta}{\eta R^2} \mathcal J(\tan \theta) + \omega,
\end{eqnarray}
where  $\mathcal J(\mu)$ is the reciprocal function of $\mu(J)$.

For frictionless suspensions, experiments suggest that the thickness $h= 50 d$ and that the rheology  is monotonous (velocity-strengthening) for all $J$ (Fig. \ref{final}\textit{A}, grey curve). Following measurements, we thus take 
\begin{eqnarray}\label{eqrheonofric}
\mu(J)  & = &  \mu_0 +(J/J_0)^\alpha 
\end{eqnarray}
with $\mu_0=0.104$, $J_0=2.27\times 10^{-2}$ and $\alpha=0.37$. 

For frictional suspensions, experiments suggest that the thickness $h= 25 d$ and that the rheology  is non-monotonous with a minimum at a critical viscous number $J_c$  (Fig. \ref{Fig8}\textit{A}, green curve). Following measurements, we take
\begin{eqnarray}
\mu(J)  & = &  \mu_{\rm stop} + ((J-J_c)/J_0)^\beta \quad {\rm for}\quad  J>J_c \label{eqrheofric1} \\
J & = & 0 \qquad {\rm for} \qquad \mu<\mu_{\rm stop} \label{eqrheofric2}
\end{eqnarray}
with $\mu_{\rm stop}=0.389$, $J_c=10^{-6}$, $J_0=8.69\times 10^{-3}$ and $\beta=0.7$. When the flow is at rest ($J=0$), the condition of start is $\tan \theta=\mu_{\rm{start}}=0.51$. 

In the following, we solved numerically Eqs. [\ref{eqdiff},\ref{eqrheofric1}--\ref{eqrheonofric}] for two cases: (i) a steady rotating drum with a constant rotation rate $\omega$, (ii) a transient relaxation ($\omega=0$) after a steady regime of constant $\omega$.  

\subsection*{(i) Steady regimes}

Fig. \ref{Fig8}\textit{B} shows the angle of avalanche obtained numerically from the model for various steady rotation rates within the experimental range.  For the frictionless rheology, one recovers that stationary values of the avalanche angle are possible for all values of rotation rates. By contrast, for the frictional rheology, stationary values of $\theta$ are possible only above a critical rotational rate $\omega_c$ corresponding to the critical viscous number $J_c$. Below this critical rotation rate, the avalanche angle exhibits the characteristic saw-tooth shape observed experimentally (compare Fig. \ref{Fig7}\textit{B} with Fig. 3B of the main text). In the model, the amplitude of hysteresis $\Delta \theta$ is only set by the difference $\mu_{\rm{start}} - \mu_{\rm stop}$ and is independent of the rotation rate, as in the experiments.  Note that in both the frictional and frictionless cases, the stability of the steady solution is a direct consequence of the velocity-strengthening part of the $\mu(J)$ curve, as a linear stability analysis of Eqs. [\ref{eqdiff},\ref{eqrheofric1}--\ref{eqrheonofric}] shows.

\subsection*{(ii) Relaxations}
Fig.\ \ref{Fig8}\textit{C} shows the relaxation dynamics of the avalanche angle obtained numerically from the model when the drum is stopped after a continuous rotation. Again, the main features of the experimental observations are recovered. For the frictionless rheology, the avalanche angle relaxes asymptotically and smoothly towards equilibrium when $t\to\infty$. By contrast, for the frictional rheology, the avalanche angle relaxes in finite time with a discontinuity of the time-derivative of the angle when the flow stops. These dynamics can be understood by writing Eqs. [\ref{eqdiff},\ref{eqrheofric1}--\ref{eqrheonofric}]  in the limit of small angle variations $\theta - \theta_0$ with $\theta - \theta_0 \ll \theta_0$. 

For frictionless grains:
\begin{eqnarray}
\frac{d\left(\theta - \theta_0\right)}{dt} &\approx& - \frac{J_0 (1 + \mu_0^2) ^{1/\alpha}}{\tau} \  \left(\theta - \theta_0\right)^{1/\alpha}\label{limnofric} \\ 
\nonumber
\end{eqnarray}

For frictional grains:
\begin{eqnarray}\label{limfric}
\frac{d\left(\theta - \theta_{stop}\right)}{dt} &\approx& - \frac{1}{\tau}\left[ J_c + J_0(1 + \mu_{stop}^2) ^{1/\beta}\ \left(\theta - \theta_{stop}\right)^{1/\beta} \right] \nonumber   \\ 
\end{eqnarray}

With $\tau^{-1} = \frac{h^3 \Delta\rho g \phi \cos\theta_0}{\eta R^2} $, $\theta_0=\arctan(\mu_0)$ and $\theta_{\rm stop}=\arctan(\mu_{\rm stop})$. For frictionless grains, Eq. [\ref{limnofric}] implies that $\theta - \theta_0$ is strictly positive and tends to zero at infinity with a power law $\theta - \theta_0\sim t^{-{\alpha/(1-\alpha)}}$. 
For frictional grains,  Eq. [\ref{limfric}] implies that $\theta - \theta_{stop}\sim  t^{-{\beta/(1-\beta)}}$ when $J_0(1 + \mu_{stop}^2)\left(\theta - \theta_{stop}\right)^{1/\beta}\gg J_c$ (initial times) and $\theta - \theta_{stop}\sim  J_c(t_f-t)/\tau$ when $J_0(1 + \mu_{stop}^2)\left(\theta - \theta_{stop}\right)^{1/\beta}\ll J_c$. Therefore, relaxation occurs in a finite time $t_f$  with a finite $\dot\theta=-J_c/\tau = -\omega_c$. Note that with $\alpha=0.37$ and $\beta=0.7$, the power law of the relaxation for the frictionless grains is smaller than for the frictional grains ($\beta/(1-\beta) > \alpha/(1-\alpha)>0$). This implies that the duration of the relaxation is shorter for the frictional grains than for the frictionless grains, as observed experimentally.

%

\bibliography{PERRIN_arXiv}

\end{document}